\documentclass[preprintnumbers,amsmath,amssymb,floatfix,11pt,prd,onecolumn,showpacs,
superscriptaddress,nofootinbib]{revtex4}
\usepackage{graphicx}
\usepackage{epsfig}
\usepackage{bm}
\usepackage{amsfonts}

\def\ben{\begin{equation}}
\def\een{\end{equation}}

 \def\bd{\begin{document}} \def\ed{\end{document}}
\def\ds{\documentstyle} \let\fr=\frac \let\bl=\bigl \let\br=\bigr
\let\Br=\Bigr \let\Bl=\Bigl
\let\bm=\bibitem
\let\na=\nabla
\let\pa=\partial \let\ov=\overline
\newcommand{\be}{\begin{equation}}
\newcommand{\ee}{\end{equation}}
\def\ba{\begin{array}}
\def\ea{\end{array}}
\def\ft#1#2{{\textstyle{\frac{\scriptstyle #1}{\scriptstyle #2} } }}
\def\fft#1#2{{\frac{#1}{#2}}}
\def\del{\partial}
\def\vp{\varphi}
\def\sst#1{{\scriptscriptstyle #1}}
\def\oneone{\rlap 1\mkern4mu{\rm l}}
\def\td{\tilde}
\def\wtd{\widetilde}
\def\ie{{\it i.e.\ }}
\def\dalemb#1#2{{\vbox{\hrule height .#2pt
        \hbox{\vrule width.#2pt height#1pt \kern#1pt
                \vrule width.#2pt}
        \hrule height.#2pt}}}
\def\square{\mathord{\dalemb{6.8}{7}\hbox{\hskip1pt}}}
\newcommand{\ho}[1]{$\, ^{#1}$}
\newcommand{\hoch}[1]{$\, ^{#1}$}
\newcommand{\bea}{\setlength\arraycolsep{2pt} \begin{eqnarray}}
\newcommand{\eea}{\end{eqnarray}}
\newcommand{\ra}{\rightarrow}
\newcommand{\lra}{\longrightarrow}
\newcommand{\Lra}{\Leftrightarrow}
\newcommand{\bp}{\tilde \beta^\prime}
\newcommand{\tr}{{\rm tr} }
\newcommand{\Tr}{{\rm Tr} }
\def\0{{\sst{(0)}}}
\def\1{{\sst{(1)}}}
\def\2{{\sst{(2)}}}
\def\3{{\sst{(3)}}}
\def\4{{\sst{(4)}}}
\def\5{{\sst{(5)}}}
\def\6{{\sst{(6)}}}
\def\7{{\sst{(7)}}}
\def\8{{\sst{(8)}}}
\def\m{{\sst{(m)}}}
\def\n{{\sst{(n)}}}
\def\cA{{{\cal A}}}
\def\cB{{{\cal B}}}
\def\cF{{{\cal F}}}
\def\cG{{{\cal G}}}
\def\cH{{{\cal H}}}
\def\tV{\widetilde V}
\def\tW{\widetilde W}
\def\tH{\widetilde H}
\def\tE{\widetilde E}
\def\tF{\widetilde F}
\def\tA{\widetilde A}
\def\im{{{\rm i}}}
\def\tY{{{\wtd Y}}}
\def\ep{{\epsilon}}
\def\vep{{\varepsilon}}
\def\bD{{{\bar D}}}
\def\R{{{\mathbb R}}}
\def\C{{{\mathbb C}}}
\def\H{{{\mathbb H}}}
\def\CP{{{\mathbb C}{\mathbb P}}}
\def\RP{{{\mathbb R}{\mathbb P}}}
\def\Z{{{\mathbb Z}}}
\def\bA{{{\mathbb A}}}
\def\bB{{{\mathbb B}}}
\def\bC{{{\mathbb C}}}
\def\bD{{{\mathbb D}}}
\def\bE{{{\mathbb E}}}
\def\bZ{{{\mathbb Z}}}
\def\Re{{{\frak{Re}}}}
\def\Im{{{\frak{Im}}}}
\def\cosec{{\,\hbox{cosec}\,}}
\def\Gm{{\Gamma_{\!\! -}}}
\def\Gp{{\Gamma_{\!\! +}}}
\def\stan{{standard }}
\def\nonstan{{supernumerary }}
\def\p{{\partial}}
\def\kdel#1{{\fft{\del}{\del#1}}}

\def\bog{{Bogomolny}}
\def\om{{\omega}}

\newcommand{\nnr}{\nonumber \\}
\newcommand{\pd}{\partial}
\newcommand{\ud}{\textrm{d}}
\newcommand{\dTH}{T^{\prime \, 0}_\textrm{H}}
\newcommand{\dOi}{\Omega^{\prime \, 0}_i}
\newcommand{\bx}{{\bf x}}
\begin{document}

\title{Reconstruction of Some Cosmological Models in $f(R,T)$ Cosmology}

\author{\textbf{Mubasher Jamil}}
\email{mjamil@camp.nust.edu.pk} \affiliation{Eurasian International
Center for Theoretical Physics, Eurasian National University, Astana
010008, Kazakhstan}\affiliation{Center for Advanced Mathematics and
Physics, National University of Sciences and Technology, H-12,
Islamabad, Pakistan}

\author{\textbf{D. Momeni}}
\email{d.momeni@yahoo.com}
 \affiliation{Eurasian International Center for Theoretical Physics, Eurasian National University, Astana 010008, Kazakhstan}

\author{\textbf{Muhammad Raza}}
\email{mreza06@gmail.com} \affiliation{Department of Mathematics,
COMSATS Institute of Information Technology (CIIT), Sahiwal campus,
Pakistan}

\author{\textbf{Ratbay Myrzakulov}}
\email{rmyrzakulov@csufresno.edu}\affiliation{Eurasian International Center
for Theoretical Physics, Eurasian National University, Astana
010008, Kazakhstan}
\begin{abstract}
\vspace*{1.5cm} \centerline{\bf Abstract} \vspace*{1cm}

In this paper, we reconstruct  cosmological models in the framework
of $f(R,T)$ gravity, where $R$ is the Ricci scalar and $T$ is the
trace of the stress-energy tensor. We show that the dust fluid
reproduces $\Lambda $CDM, phantom-non-phantom era and the phantom
cosmology. Further, we reconstruct  different cosmological models
including, Chaplygin gas, scalar field with
some specific forms of $f(R,T)$. Our numerical simulation for Hubble parameter shows good agreement with the BAO observational data for low redshifts $z<2$.
\end{abstract}
\pacs{04.20.Cv, 04.50.Kd, 98.80.Jk, 98.80.Bp} \maketitle
\newpage

\section{Introduction}

From the cosmological observational data \cite{riess,riess2,riess3,riess4}, it is now
well established that the present observable Universe is undergoing
an accelerating expansion. While the source driving this cosmic
acceleration is known as `dark energy' its origin has not been
well understood yet due to absence of a consistent theory of quantum
gravity. This acceleration is driven by the negative pressure of the
dark energy. The `cosmological constant' is the most simple and
natural candidate for explaining cosmic acceleration but it faces
serious problems of fine-tuning and large mismatch between theory
and observations \cite{reviews,r2,r3}. Hence there has been significant
development in the construction of dark energy models by modifying
the geometrical part of the Einstein-Hilbert action. This
phenomenological approach is called as the Modified Gravity which
can successfully explain the rotation curves of galaxies, the motion
of galaxy clusters, the Bullet Cluster, and cosmological
observations without the use of dark matter or Einstein's
cosmological constant \cite{luca,l2,l3,l4,l5,l6,l7,l8,l9,mofat,m2,m3}. The $f(R)$ theories can
produce cosmic inflation, mimic behavior of dark matter and current
cosmic acceleration, being compatible with the observational data
\cite{NO,noji,carr,Star} (also see a recent review \cite{review1} on
$f(R)$ gravity and its cosmological implications ).

In a recent paper \cite{sergei2011}, the authors considered a
generalized gravity model $f(R,T)$, with $T$ being the trace of
stress-energy tensor, manifesting a coupling between matter and
geometry. By choosing different functional forms of $f$, they solved
the dynamical equations of astrophysical and cosmological interest.
In the present work, we study the same model by taking different
kinds of  the energy sources.  Reconstruction of the cosmological
models for this theory have been ivestigated by several authors
\cite{1111.4275,1110.5756,1110.1049,1109.2928,1107.3887}.

The paper is organized as follows. In section II we present the general
 action and the equation of the motion for $f(R,T)$. In section III
  we reconstruct the cosmological models from the dust fluid.
   In section IV we generalize the dust fluid models to the general fluid
   with EoS $p=\omega \rho$.  Section V is devoted to
  discussion and conclusions.

\section{Field equations in $f(R,T)$ gravity}

The action of $f(R,T)$ gravity is of the form \cite{sergei2011}
\begin{equation}\label{1d}
S=\frac{1}{16\pi}\int d^4x\sqrt{-g}f(R,T)d^4x +\int L_m\sqrt{-g}d^4x,
\end{equation}
where $f(R,T)$ is an arbitrary function of the scalar curvature
$R=R^\mu_\mu$ and the trace $T=T^\mu_\mu$ of the energy-momentum
tensor $T_{\mu\nu}$. We define the Lagrangian density for matter field $L_m$
by
\begin{equation}\label{2d}
T_{\mu\nu}=-\frac{2}{\sqrt{-g}}\frac{\delta(\sqrt{-g}L_m)}{\delta
g^{\mu\nu}}.
\end{equation}
The equation of motion (EOM) is obtained by varying the action (\ref{1d}) with respect to $g^{\mu\nu}$ \cite{sergei2011}
\begin{equation}\label{eom}
f_{R}R_{\mu\nu}-\frac{1}{2}Rg_{\mu\nu}+(g_{\mu\nu}\Box-\nabla_{\mu}\nabla_{\nu})f_{R}=8\pi
T_{\mu\nu}-f_{T}T_{\mu\nu}-f_{T}\Theta_{\mu\nu},
\end{equation}
where $f_R(R,T)=\frac{\partial f(R,T)}{\partial R}$ and $f_T(R,T)=\frac{\partial f(R,T)}{\partial T}$. $\nabla_\mu$ is the operator for covariant derivative and box operator (or d' Alembert operator) $\Box$ is defined via  $$\Box\equiv\frac{1}{\sqrt{-g}}\partial_\mu(\sqrt{-g}g^{\mu\nu}\partial_\nu),\ \ \ \Theta_{\mu\nu}\equiv g^{\alpha\beta}\frac{\delta
T_{\alpha\beta}}{\delta g^{\mu\nu}}.$$
Performing a contraction of indices in (\ref{eom}), we obtain
\begin{eqnarray}\label{eq}
R f_R+3\Box f_R-2f=8\pi T-T f_T-\Theta f_T.
\end{eqnarray}
Here $\Theta\equiv g^{\mu\nu}\Theta_{\mu\nu}$. We will use (\ref{eq}) to reconstruct $f(R,T)$ for different kinds of  matter.

\section{Reconstruction of $f(R,T)$ Using Dust}
 
We consider the metric of a flat Friedmann-Robertson-Walker (FRW)
spacetime
\begin{equation}\label{metric}
g_{\mu\nu}=diag(1,-a^2(t),-a^2(t),-a^2(t)).
\end{equation}
Following the conservation of the energy momentum tensor
$T^{\mu\nu}_{;\mu}=0$ for metric (\ref{metric}) we obtain
\begin{equation}\label{tdot}
\dot{T}=-3H T.
\end{equation}
The $00-$component of (\ref{eom}) reads
\begin{equation}\label{8d}
f_{R}\Big(\frac{2}{3}R-\ddot{R}\Big)+\frac{1}{6}f=\frac{2}{3}T(8\pi+f_T)+\dot{R}^
2f_{RR}+9H^2T^2
f_{TT}-6HT\dot{R}f_{RT}-3T\Big(\frac{R}{6}-5H^2\Big)f_T.
\end{equation}
In (\ref{8d}), if we know the functional form of the Hubble
parameter $H$ as a function of the two independent variables $R$ and
$T$, then reconstruction of the model $f(R,T)$ is straightforward. But
always, we have a physical intuition about the form of $H(R,T)$. We
clarify it here more. In a FRW model, the form of the Ricci scalar
is $R=6(\dot{H}+2H^2)$. Any cosmological era has a specific matter
fields, radiation, phantom field or a mixed of them. For example in
 $\Lambda $CDM model,
we have $H^2=H_0^2+\frac{\kappa^2 \rho_0 a^{-3}}{3}$. It is easy to
see that we can write $R$ in terms of $H$ as
$H^2=\frac{R}{3}-3 H_0^2$. Thus usually, we can write $H^2=H^2(R)$.
Thus (11) can be integrated easily to give the functional form
$f(R,T)$. Also, since $\dot{T}=-3H(R)T$,
$\ddot{T}=-3T(\frac{R}{6}-5H^2)$ thus all terms of the (\ref{tdot}) are
functions of $R$, $T$. For term as $\ddot{R}$ we can write it as
$\ddot{R}=\frac{\dot{H}}{\frac{d
H}{dR}}=\frac{\frac{R}{6}-2H(R)^2}{\frac{d H}{dR}}$. Mathematically,
(\ref{8d}) is a second order partial differential equation for
$f(R,T)$.

\subsection{Cosmological Implications}

We now discuss the solutions of (\ref{8d}) relevant in cosmological
context. In the coming sections we will show that, any cosmological
epoch (radiation, matter, dark energy dominated eras) can be
constructed in a model of $f(R,T)$ only with dust fluid as the
source. We mention here that the form of the action $f(R,T)$ is not
unique. For this reason we can continue our investigation of the
cosmological reconstruction for another model with
$f(R,T)=f_1(R)+f_2(T)$ \cite{sergei2011}. When we set $f_2(T)=0$, we
recover the $f(R)$ theory which has been discussed in
\cite{sergei2009}. Adopting the technique of the reconstruction of
$f(R,T)$ models, we have the following different models reconstructed from the dust fluid.

\subsection{$\Lambda$CDM model}

In Einstein gravity, the Hubble parameter for a flat FRW model with
real matter field describes the $\Lambda $CDM model by
$H^2=H_0^2+\frac{\kappa^2 }{3}\rho_0 a^{-3}$, in units
$\kappa^2=8\pi G$ $c=1$. Since
\begin{eqnarray}
H^2&=&\frac{R}{3}-3 H_0^2,\\
\dot{R}&=&-9\Big(\frac{R}{3}-3 H_0^2\Big)^{3/2},\\
\ddot{R}&=&\frac{9}{2}(R-12H_0^2)(R-9H_0^2),
\end{eqnarray}
then  the field equation (\ref{8d}) converts to the following form
\begin{eqnarray}\label{eqa}
\Big[\frac{2}{3}R-\frac{9}{2}\Big(R-12H_0^2\Big)\Big(R-9H_0^2\Big)\Big]f_{R}+
\frac{1}{6}f=\frac{2}{3}T(8\pi+f_T)+81\Big(\frac{R}{3}-3
H_0^2\Big)^{3}f_{RR}\\\nonumber+9\Big(\frac{R}{3}-3 H_0^2\Big)T^2
f_{TT}+54\Big(\frac{R}{3}-3
H_0^2\Big)^2Tf_{RT}-3T\Big(\frac{R}{6}-5H^2\Big)f_T.
\end{eqnarray}
The solutions of (\ref{eqa}) can be obtained under two special cases
as:
\begin{itemize}
\item $f(R,T)=F(R)$ and $F(R)=AR+B$ where $A$, $B$ and $T$ are
constants. Equation (15) gives $B=32\pi T$ and $A=0$.

\item $f(R,T)=G(T)$ and $G(T)=CT+D$ where $C$, $D$ and $R$ are
constants. Equation (15) gives
$C=\frac{-16/3\pi}{\frac{1}{2}+\frac{9}{2}R+9H_0^2}$ and $D=0$.
\end{itemize}

\subsection{$f(R,T)$  reproducing the system with phantom and
non-phantom matter}

In Einstein gravity, a Universe filled with a mixture of the phantom
and non-phantom components obey from the following expression for H,
$H^2=\frac{\kappa^2}{3}(\rho_q a^{-c_1}+\rho_p a^{c_1})$. Here the
set of the variables $\rho_q$ (energy density of non-phantom matter)
$ \rho_p,$ (energy density of phantom matter) $c_1$ are the
parameters of the model. In the early Universe when the scale factor
was small, the first term in $\rho_q a^{-c_1}$ dominates and
\emph{it behaves as the Universe described by the Einstein gravity
with a matter whose EoS parameter is $w=-1+c_1/3>-1$} which means
that the matter field is non-phantom like. But for present era we
have $w=-1-c_1/3<-1$ which means we live in a phantom epoch of the
Universe.

As a special case, $c_1=4$ which is a special form discussed
\cite{sergei2009} we have
\begin{eqnarray}
H^2&=&\frac{R}{3}-3 H_0^2,\\
\dot{R}&=&2\frac{\sqrt{\frac{a}{R}+bR}(\frac{R}{6}-2 H^2)}{b-\frac{a}{R^2}},\\
\ddot{R}&=&\{2\frac{\sqrt{\frac{a}{R}+bR}(\frac{R}{6}-2
H^2)}{b-\frac{a}{R^2}}\}\frac{d}{dR}\Big[2\frac{\sqrt{\frac{a}{R}+bR}(\frac{R}{6}-2
H^2)}{b-\frac{a}{R^2}}\Big].
\end{eqnarray}
Thus (\ref{8d}) converts to the following form
\begin{eqnarray}\label{phantom}
&&f_{R}(\frac{2}{3}R-\{2\frac{\sqrt{\frac{a}{R}+bR}(\frac{R}{6}-2
H^2)}{b-\frac{a}{R^2}}\}\frac{d}{dR}[2\frac{\sqrt{\frac{a}{R}+bR}(\frac{R}{6}-2
H^2)}{b-\frac{a}{R^2}}])\\&&\nonumber+\frac{1}{6}f=\frac{2}{3}T(8\pi+f_T)
+(2\frac{\sqrt{\frac{a}{R}+bR}(\frac{R}{6}-2
H^2)}{b-\frac{a}{R^2}})^2f_{RR}+9(\frac{R}{3}-3 H_0^2)T^2
f_{TT}\\&&\nonumber-6\sqrt{\frac{R}{3}-3
H_0^2}T\{2\frac{\sqrt{\frac{a}{R}+bR}(\frac{R}{6}-2
H^2)}{b-\frac{a}{R^2}}\}f_{RT}-3T(\frac{R}{6}-5H^2)f_T.
\end{eqnarray}
We solve equation (\ref{phantom}) using an ansatz $f(R,T)=F(T)$. The solution
of (\ref{phantom}) is
\begin{eqnarray}
F(T)=C_1T^{k_1}+C_2T^{k_2}+T-\frac{32\pi T}{3+27R-270H_0^2},
\end{eqnarray}
where $C_1$ and $C_2$ are two constants of integration and
\begin{eqnarray}
k_1=\frac{1}{36}\frac{-4-9R+108H_0^2+\sqrt{16+144R-1512H_0^2+81R^2-1944RH_0^2+11664H_0^4}}{R-9H_0^2},
\end{eqnarray}
\begin{eqnarray}
k_2=\frac{-1}{36}\frac{4+9R-108H_0^2+\sqrt{16+144R-1512H_0^2+81R^2-1944RH_0^2+11664H_0^4}}{R-9H_0^2}.
\end{eqnarray}
If we choose $f(R,T)=C_3$ a pure constant, then Eq.  (\ref{phantom}) gives
$C_3=32\pi T$ where $T$ is also a constant. Furthermore if
$f(R,T)=F(R)$ then Eq.  (\ref{phantom}) cannot be solved analytically or
numerically.

\subsection{de Sitter Universe in $f(R,T)$ Gravity}

If we live in a Universe filled by dust and the scale factor
increases exponentially with time as $a(t)=a_0 e^{H_0 t}$, then
Hubble parameter is constant. Such model which has been proposed
firstly by de Sitter, is called the de Sitter Universe. In this
solution, it is assumed that the only source of matter filling the
Universe is dust. Thus it is contained in our treatment on dust
formation. For a de Sitter Universe in it's static patch  we've
\begin{eqnarray}
H=H_{0},\\R=12 H_{0}^{2},\\\dot{R}=\ddot{R}=0,
\end{eqnarray}
Eq. (\ref{8d}) reduces to
\begin{equation}
\frac{2}{3}Rf_{R}+\frac{1}{6}f=\frac{2}{3}T(8\pi+f_T)+9H_0^{2}T^2
f_{TT}-3T\Big(\frac{R}{6}-5H_0^{2}\Big)f_T.\label{desitter}
\end{equation}
Solution of equation (\ref{desitter}) is
\begin{equation}
f(R,T)= F_1 T^{k_3}+ F_2T^{k_4}-\frac{32\pi T}{3+54H_0^{2}},
\end{equation}
where $F_1$ and $F_2$ are two arbitrary constants and
\begin{equation}
k_3=\frac{1}{54}\Big(\frac{-2+\sqrt{4+54H_0^{2}}}{H_0^{2}}\Big),
\end{equation}
\begin{equation}
k_4=\frac{-1}{54}\Big(\frac{2+\sqrt{4+54H_0^{2}}}{H_0^{2}}\Big).
\end{equation}

\subsection{Einstein Static Universe in $f(R,T)$ Model}

In this case since $H=0$, thus we have
$\dot{H}=R=\dot{R}=\ddot{R}=0$. Thus we have
\begin{equation}
\frac{1}{6}f=\frac{2}{3}T(8\pi+f_T).
\end{equation}
The solution for this purely $T$ dependence equation is
\begin{equation}
f(T)=-\frac{32 \pi T}{3}+C_4 T^{1/4}.
\end{equation}

\section{Models for $p=\omega\rho$}

Starting from (\ref{eq}) and assuming EoS $p=\omega\rho$,  we have
\begin{eqnarray}
\Theta_{\mu\nu}=-2(1+\omega)\rho u_{\mu}u_{\nu}+p.
g_{\mu\nu}\label{teta}
\end{eqnarray}
The trace of (\ref{teta}) is given by
\begin{eqnarray}
\Theta=2\rho(\omega-1).
\end{eqnarray}
Since $T=\rho (1-3\omega)$ thus, we have
\begin{eqnarray}
\Theta=\frac{2\omega-2}{1-3\omega}T. \label{teta2}
\end{eqnarray}
Now we rewrite (\ref{eq}) in the following form
\begin{eqnarray}
R f_R+3\Box f_R-2f=8\pi T-T f_T-\frac{2\omega-2}{1-3\omega}T
f_T,\label{eq2}
\end{eqnarray}
which can be written in suitable form as
\begin{eqnarray}
R f_R+3\Box f_R-2f=8\pi T-\omega'T f_T,\label{eq3}
\end{eqnarray}
where $\omega'=-\frac{\omega+1}{1-3\omega}$. We will 
search for exact solutions of $f(R, T )$ in the following cases.

\subsection{Solutions in the form $f(R, T)=R+2 f(T )$}

Substituting this form $f(R, T)=R+2 f(T )$ in (\ref{eq3}) we obtain
\begin{eqnarray}
-2R -4 f(T )=8\pi T-2\omega'T f_T.\label{eq4}
\end{eqnarray}
There is only one possibility to get an exact solution: when
$R$ is constant.In this case, the solution for (\ref{eq4}) reads
\begin{eqnarray}
 f(T )=c_1 T^{2/\omega'}+\frac{2(2-\omega')R+16\pi T}{4 (\omega'-2)}.
\end{eqnarray}
Thus we have
\begin{eqnarray}
f(R, T)=R+2 \Big(c_1 T^{2/\omega'}+\frac{2(2-\omega')R+16\pi T}{4
(\omega'-2)}\Big). \label{sol1}
\end{eqnarray}

\subsection{Solutions in the form $f(R, T)=f_1(R)+ f_2(T )$}

Substituting this form of $f(R, T)=f_1(R)+ f_2(T )$ in (\ref{eq3})
we obtain
\begin{eqnarray}
R f'_{1}(R)+3\Box f'_{1}(R)-2f_1(R)=8\pi T-\omega'T
f'_2(T)+2f_2(T).\label{eq5}
\end{eqnarray}
Note that the left and right hand sides of (\ref{eq5}) are functions of $R$ and $T$ respectively. Thus solving the $T$ dependent part of (\ref{eq5}), we obtain 
\begin{eqnarray}
f_2(T)=C T^{\frac{2}{\omega'}}+\frac{16\pi
T+(\omega'-2)c_1}{2(\omega'-2)}.
\end{eqnarray}
The solution for the $R$ dependent part of the (\ref{eq5}) is
complicated. Indeed we must solve the following equation
\begin{eqnarray}
R
f'_{1}(R)+3(\partial_{tt}f'_{1}(R)+3H\partial_{t}f'_{1}(R))-2f_1(R)=c_1,\label{eq6}
\end{eqnarray}
or the following equivalent form
\begin{eqnarray}
R f'_{1}(R)+3(\ddot{R}f''_{1}(R)+\dot{R}^2
f'''_{1}(R)+3H\dot{R}f''_{1}(R))-2f_1(R)&=&\frac{c_1}{6},\label{eq7}
\end{eqnarray}
One simple but interesting solution  is obtained by taking Ricci curvature to be constant. From (\ref{eq7}) we have
\begin{eqnarray}
f_{1}(R)=-\frac{c_1}{2}.
\end{eqnarray}
Thus one of the interesting models is
\begin{eqnarray}
f(R, T)=-\frac{1}{2}c_1+C T^{\frac{2}{\omega'}}+\frac{16\pi
T+(\omega'-2)c_1}{2(\omega'-2)}.\label{sol2}
\end{eqnarray}
\begin{figure}
\centering
 \includegraphics[scale=0.4] {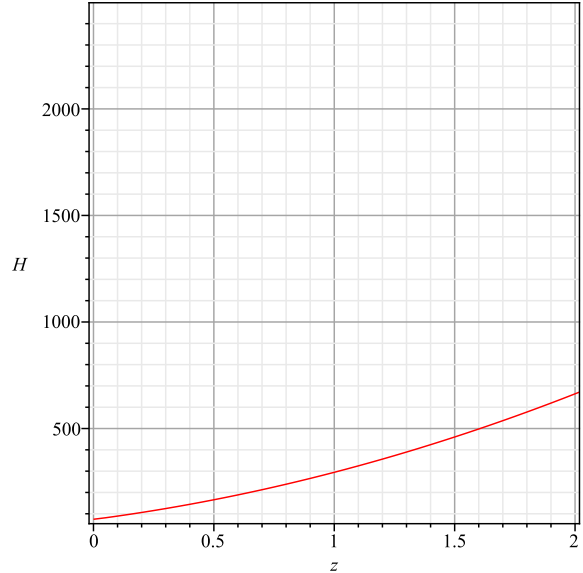}
  \caption{ Evolution of Hubble parameter (\ref{david1}) against redshift $z$. We choose $c_1=6$, $H(0)=H_0=74.2$, $\frac{dH}{dz}(z=0)=22.26$.  }
\end{figure}

\begin{figure}
\centering
 \includegraphics[scale=0.4] {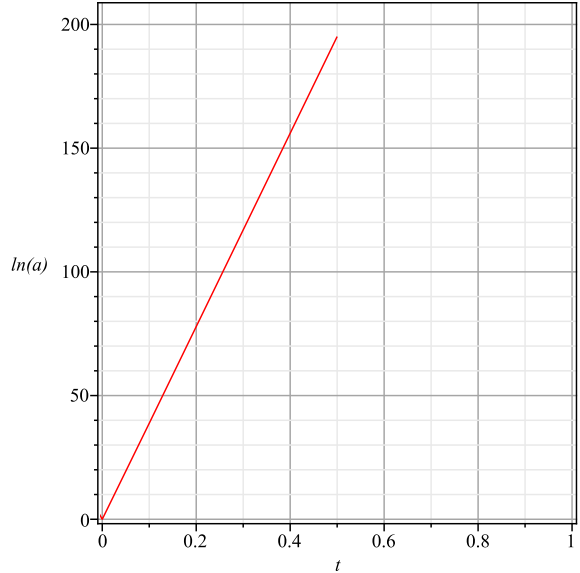}
  \caption{ Evolution of $\log (a)$ (\ref{david2}) against time $t$. We choose $c_1=6$, $a(0)=1$, $\dot a(0)=74.2$, $\ddot{a}(0)=2.73\times10^5$, $\dot{\ddot{a}}(0)=4\times10^5$ }
\end{figure}

For this model, we can study the evolution of the Hubble parameter and the scale factor. For this purpose, we assume $f_1(R)=R^2$ and rewrite (\ref{eq7}) in terms of the Hubble parameter $H(z)$ (and using $dt=-\frac{1}{1+z}\frac{dz}{H(z)}$) and the scale factor $a(t)$:
\begin{eqnarray}
18 H\frac{dH}{dz}-6(1+z)\Big( \frac{dH}{dz} \Big)^2-6(1+z)H\frac{d^2H}{dz^2}+3H\Big(12H^2-6H(1+z)\frac{dH}{dz}\Big)&=&\frac{c_1}{6}.\label{david1}\\
a^2\frac{d^4a}{dt^4}-5\Big(\frac{da}{dt}\Big)^2\frac{d^2a}{dt^2}+a\Big( \frac{d^2a}{dt^2} \Big)^2
+3a\frac{da}{dt}\frac{d^3}{dt^3}&=&\frac{c_1}{6}a^3.\label{david2}
\end{eqnarray} 
The two equations (\ref{david1}) and (\ref{david2}) are numerically solved and the result is shown in figures 1 and 2. The initial conditions are determined using the definitions of Hubble parameter and jerk parameter. Specifically $a(0)=1$, $H_0=74.2$, $\dot a(0)=H_0a_0=74.2$, $q_0=-0.67$, $j_0=-0.98$ \cite{q0}, $\ddot a(0)=-H_0^2q_0$, $\dot{\ddot{a}}(0)=-j_0H_0^3=4\times 10^5$ \cite{q0}. Note that our model is compatible with the BAO data \cite{bao} for the Hubble parameter.

\subsection{Reconstruction Using Chaplygin Gas}

The Chaplygin gas (CG)  has the equation of state $p=\frac{-A}{\rho}$. There are
some extensions of this model like the generalized Chaplygin gas but we restrict
ourselves only to the CG case. Thus we have
quantities
\begin{eqnarray}
\Theta=-2T+4\frac{A}{\rho},\label{tetarho}\\
T=\rho+3\frac{A}{\rho}.\label{rho}
\end{eqnarray}
Solving (\ref{rho}) for $\rho$ and inserting the solution in (\ref{tetarho})
and by defining a new parameter as $12A=T_0^{2}$ we have
\begin{eqnarray}
\Theta\equiv\Theta(T)=\frac{2}{3}\frac{T_0^2-3T^2-3T\sqrt{T^2-T_0^2}}{T+\sqrt{T^2-T_0^2}}.\label{t32}
\end{eqnarray}
Using (\ref{t32}) in (\ref{eq}), we have
\begin{eqnarray}
R f_R+3\Box f_R-2f=8\pi T-T
f_T-\frac{2}{3}\frac{T_0^2-3T^2-3T\sqrt{T^2-T_0^2}}{T+\sqrt{T^2-T_0^2}}
f_T.\label{t33}
\end{eqnarray}
We are searching for constant curvature solutions in which
$R=R_0\Rightarrow\Box f_R=0$. Thus solution of (\ref{t33}) is
\begin{eqnarray}
f( T)=R_0+e^{2\int \frac{dT}{T+\Theta (T)}}\Big[C+8\pi\int\frac{T dT}{g(T)(T+\Theta(T))}\Big]. \label{sol5}
\end{eqnarray}

\subsection{$f(R, T)$ Models for Scalar Field}

We know that there is an important duality between $f(R)$ models and
scalar fields \cite{review1}. The Lagrangian for a scalar field, which is minimally
coupled to the background reads
\begin{eqnarray}
L=-\frac{1}{2}\omega\phi_{,\mu}\phi^{,\mu}.
\end{eqnarray}
Here $\omega$ is a free parameter. The corresponding expression for
stress-energy tensor is
\begin{eqnarray}
T_{\mu\nu}=-\frac{1}{2}\omega(\phi_{,\mu}\phi_{,\mu}-\frac{1}{2}g_{\mu\nu}\phi_{,\alpha}\phi^{,\alpha}).
\end{eqnarray}
Now, the expressions for $\Theta$ and $T$ read
\begin{eqnarray}
\Theta=-3\omega\phi_{;\alpha}\phi^{;\alpha},\label{tetasf}\\
T=\frac{1}{2}\omega\phi_{;\alpha}\phi^{;\alpha}.\label{tsf}
\end{eqnarray}
Eliminating the term $\phi_{;\alpha}\phi^{;\alpha}$ from
(\ref{tetasf}) and (\ref{tsf}) we have
\begin{eqnarray}
\Theta=-6T.\label{tetat}
\end{eqnarray}
We can rewrite (\ref{eq}) using (\ref{tetat})
\begin{eqnarray}
R f_R+3\Box f_R-2f=8\pi T+5 T f_T.\label{eqsf}
\end{eqnarray}
Again, we limit ourselves to constant curvature solutions in which
$R=R_0$ (choosing $f(R,T)\longrightarrow R+f(T)$). Then, the general solution for (\ref{eqsf}) is
\begin{eqnarray}
f(R,
T)=R-\frac{4\pi}{7}T+\frac{C}{T^{\frac{2}{5}}}.\label{solsf}
\end{eqnarray}

\section{Conclusion}

In summary, we reconstruct  cosmological models in the framework of
a newly proposed model of $f(R,T)$ gravity, where $R$ is the Ricci
scalar and $T$ is the trace of the stress-energy tensor.  We show
that the dust fluid reproduces $\Lambda $CDM, Einstein static
Universe, de Sitter Universe, phantom-non-phantom era and the
phantom cosmology. Further, we reconstruct  different cosmological
models including, Chaplygin gas, minimally coupled scalar field, with some specific forms of $f(R,T)$. We demonstrated that for one case $f(R,T)=R^2+f(T)$, we found that the behavior of Hubble parameter $H(z)$ and scale factor is compatible with the observational data of BAO for small redshift $z<2$.

\subsection*{Acknowledgment}
M. Jamil and D. Momeni would like to thank the warm hospitality of Eurasian National University, Astana, Kazakhstan where this work was completed. All authors would like to thank the anonymous referee for enlightening comments on this paper.

\end{document}